\documentclass[epj]{webofc}
\usepackage[varg]{txfonts}   
\woctitle{XLVII International Symposium on Multiparticle Dynamics}
\begin{document}
\title{Sound waves in hadronic matter}
%
%

\author{Grzegorz Wilk\inst{1}\fnsep\thanks{\email{grzegorz.wilk@ncbj.gov.pl}} \and
        Zbigniew W\l odarczyk\inst{2}\fnsep\thanks{\email{zbigniew.wlodarczyk@ujk.kielce.pl}}
}

\institute{National Centre for Nuclear Research, Ho\.za 69, 00-681 Warsaw, Poland
\and
           Institute of Physics, Jan Kochanowski University, \'Swi\c{e}tokrzyska 15, 25-406 Kielce, Poland
          }

\abstract{We argue that recent high energy CERN LHC experiments on transverse momenta distributions of produced particles provide us new, so far unnoticed and not fully appreciated, information on the underlying production processes. To this end we concentrate on the small (but persistent) log-periodic oscillations decorating the observed $p_T$ spectra and visible in the measured ratios $R = \sigma_{data}\left( p_T\right)/\sigma_{fit}\left( p_T\right)$. Because such spectra are described by quasi-power-like formulas characterised by two parameters: the power index $n$ and scale parameter $T$ (usually identified with temperature $T$), the observed log-periodic behaviour of the ratios $R$ can originate either from suitable modifications of $n$ or $T$ (or both, but such a possibility is not discussed). In the first case $n$ becomes a complex number and this can be related to scale invariance in the system, in the second the scale parameter $T$ exhibits itself  log-periodic oscillations which can be interpreted as the presence of some kind of sound waves forming in the collision system during the collision process, the wave number of which has a so-called self similar solution of the second kind. Because the first case was already widely discussed we concentrate on the second one and on its possible experimental consequences.}
\maketitle
\section{Introduction}
\label{intro}

The high energy CERN LHC experiments measure, among other things, transverse momenta distributions of produced particles \cite{QP1,QP2,QP3,QP4,QP5}. We would like to draw attention the fact that, apparently, some features of the results obtained have so far remained unnoticed or not fully appreciated. They are presented in Fig. \ref{FF1}. In panel $(a)$ one can see the typical $pp$ large $p_T$ cross section which is best described by the Tsallis quasi power-like distribution where $q$ is the so-called nonextensivity parameter and $T_0$ is a scale factor sometimes identified with the temperature (we use units with $c=k_B=h=1$).
\begin{equation}
f\left( p_T\right) = \frac{n-1}{nT_0}\left( 1 + \frac{p_T}{nT_0}\right)^{-n},\quad n = \frac{1}{q-1}, \label{TD}
\end{equation}
However, if the ratios $R = \sigma_{data}\left( p_T\right)/\sigma_{fit}\left( p_T\right)$ are plotted log-periodic oscillations are seen in the observed $p_T$ spectra, see panels $(b)-(d)$. They are rather small but they show up in the results from different experiments (panel $(b)$), at different energies (panels $(c)-(d)$ and in different systems (panels $(e)-(f)$). Panel $(e)$ presents a comparison of $p_T$ distributions for $p+p$ and $Pb+Pb$ collisions whereas panel $(f)$ compares the respective factors $R$ showing that the effect increases with the centrality of the nuclear collision \cite{WWln,RWWlnA}. In panel $(f)$ we also present a comparison with the proposed alternative to our description (using a two component, {\it soft+hard}, picture of the production process, each in the form of Eq. {\ref{TD}) \cite{SoftHard1,SoftHard2}). As can be seen it is not able to describe the observed effect in the whole region of $pT$. Because it turns out that, additionally, these oscillations cannot be erased by any reasonable change of fitting parameters, we shall assume that this is a real effect which should be investigated in detail. Note that log periodic oscillations can originate either from suitable modifications of the power index $n$ (or nonextensivity parameter $q$ in our case) or from some suitable modification of the scale parameter $T$ (or from both cases, but such possibility will not be discussed).
\begin{figure}[h]
\begin{center}
\resizebox{1.0\textwidth}{!}{%
  \includegraphics{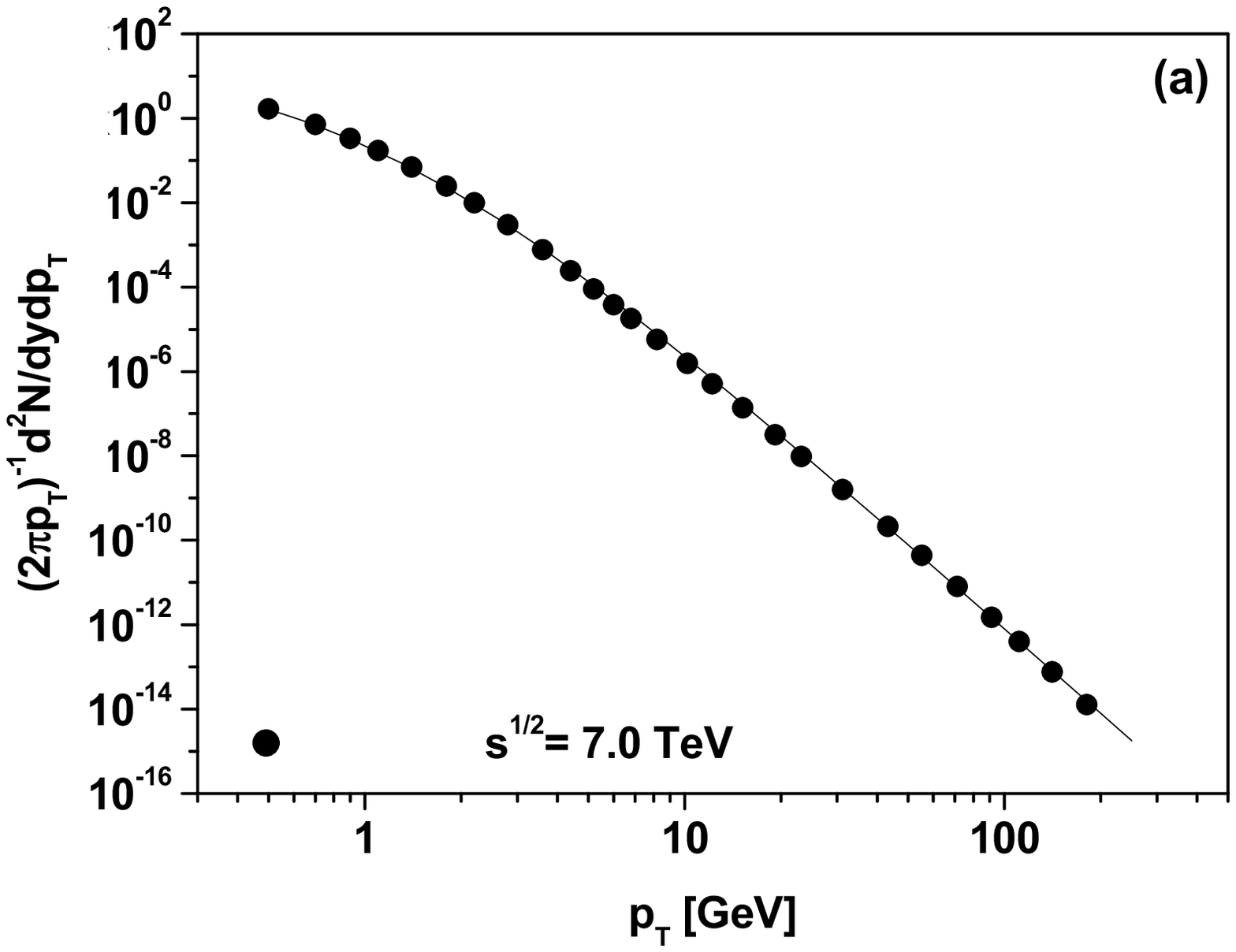}\hspace{5mm}
  \includegraphics{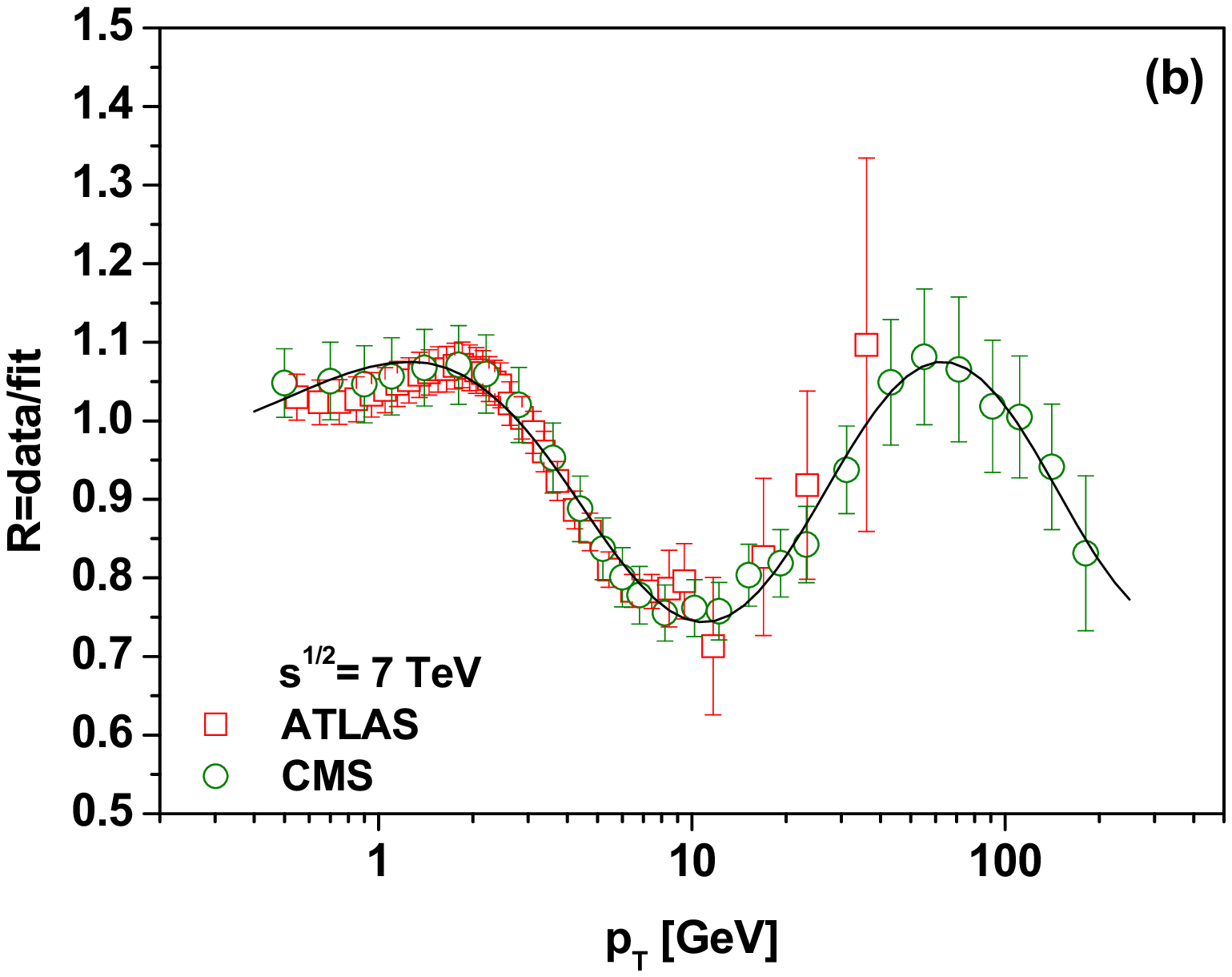}\hspace{5mm}
  \includegraphics{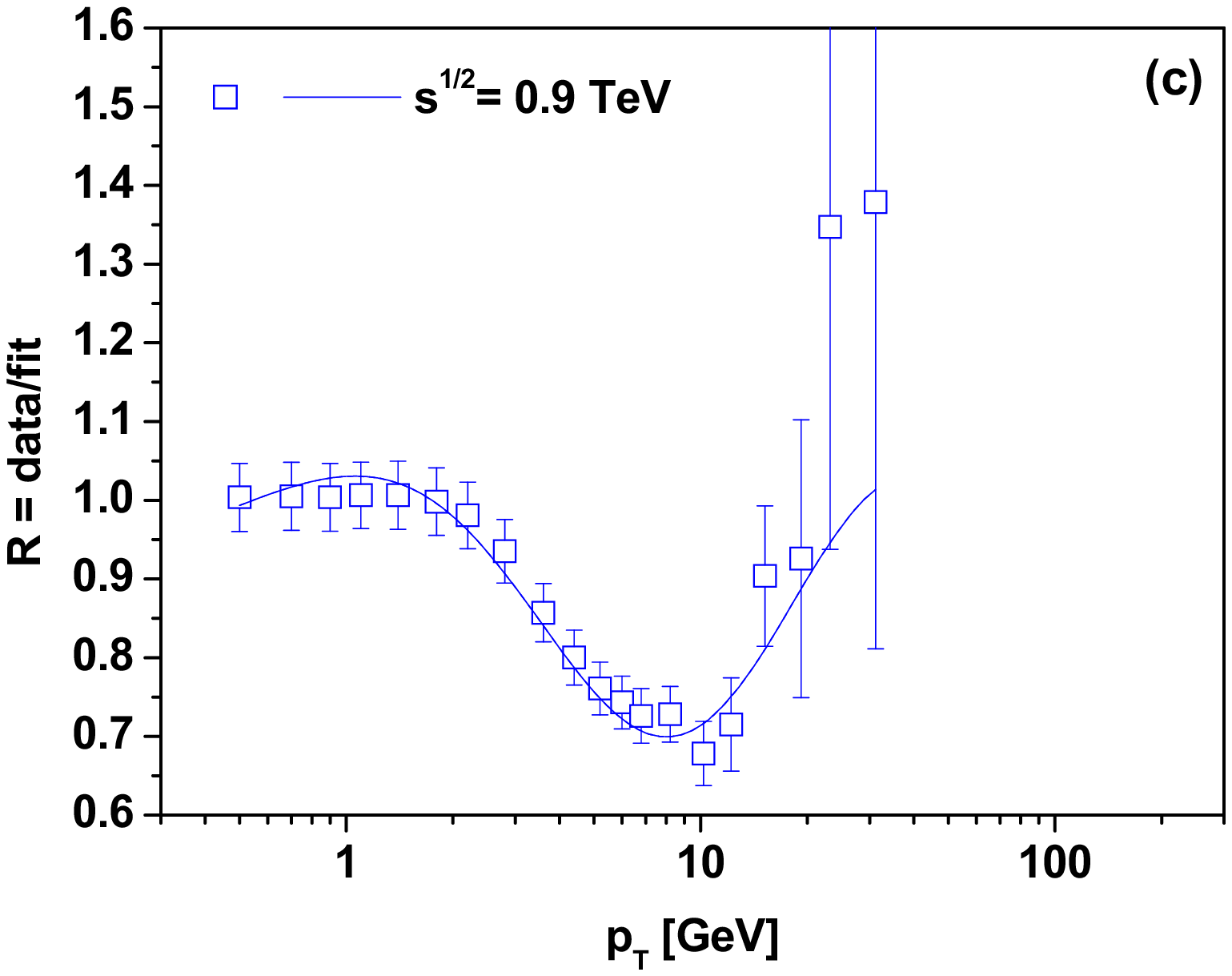}
  }
\resizebox{1.0\textwidth}{!}{%
  \includegraphics{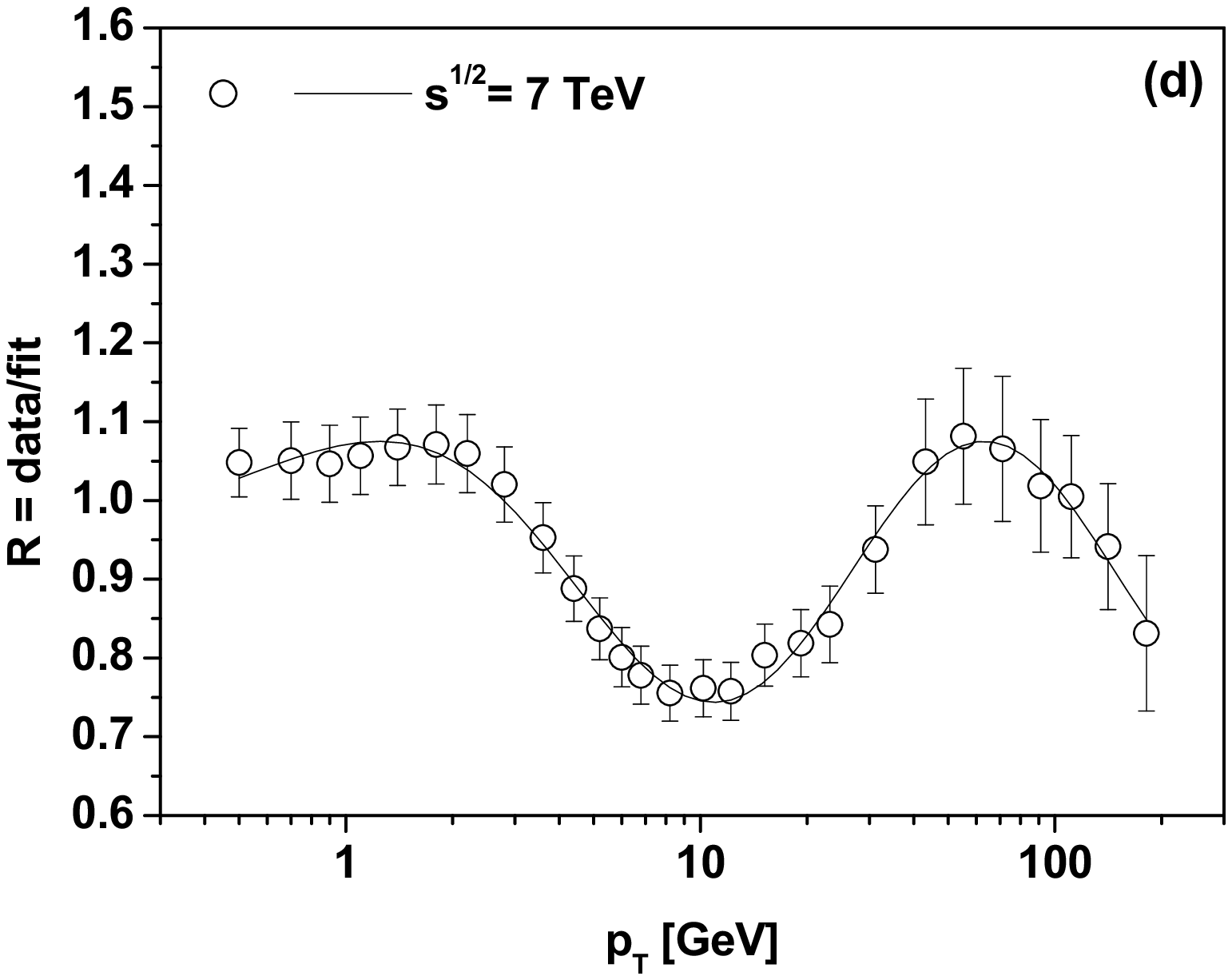}\hspace{5mm}
  \includegraphics{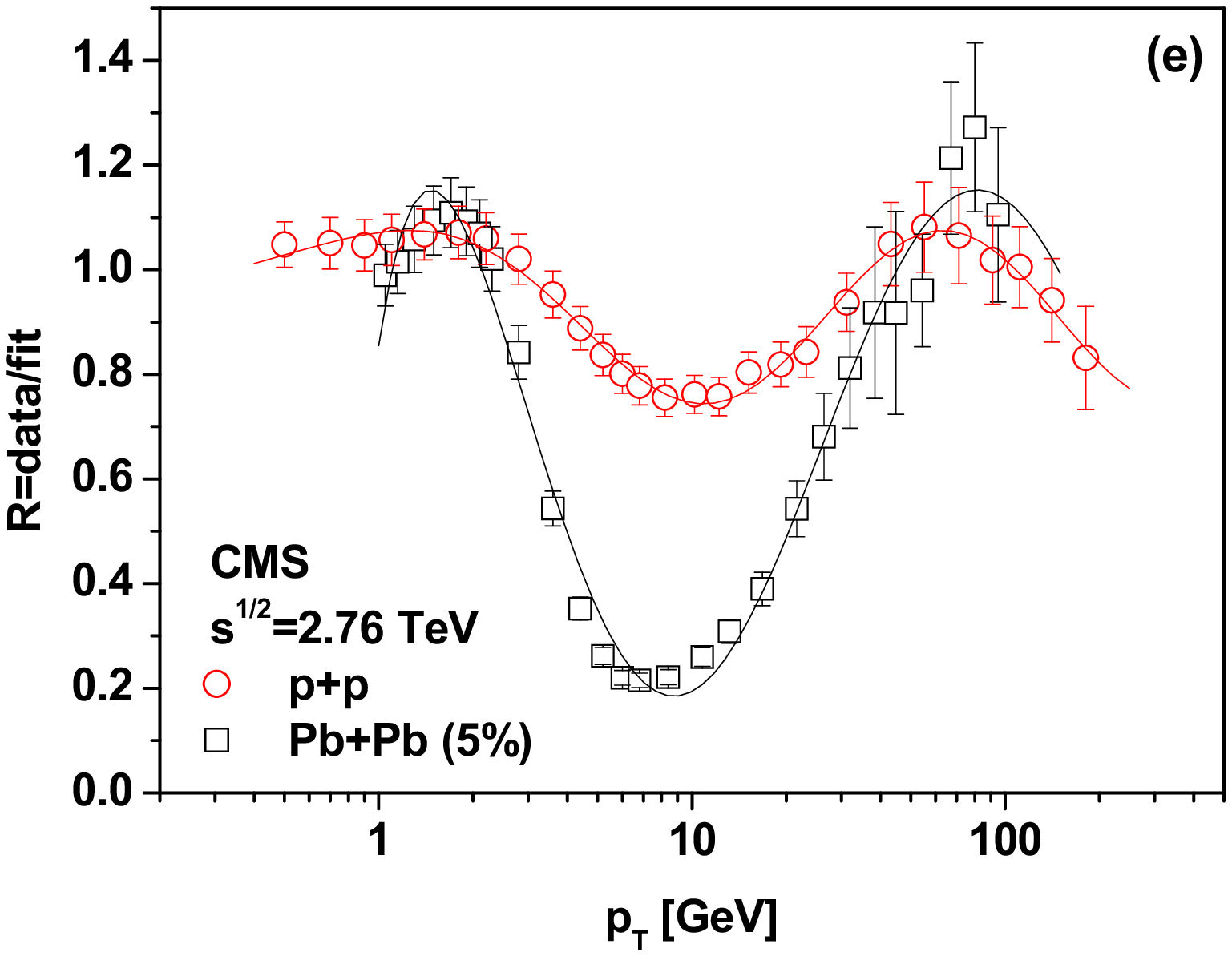}\hspace{5mm}
  \includegraphics{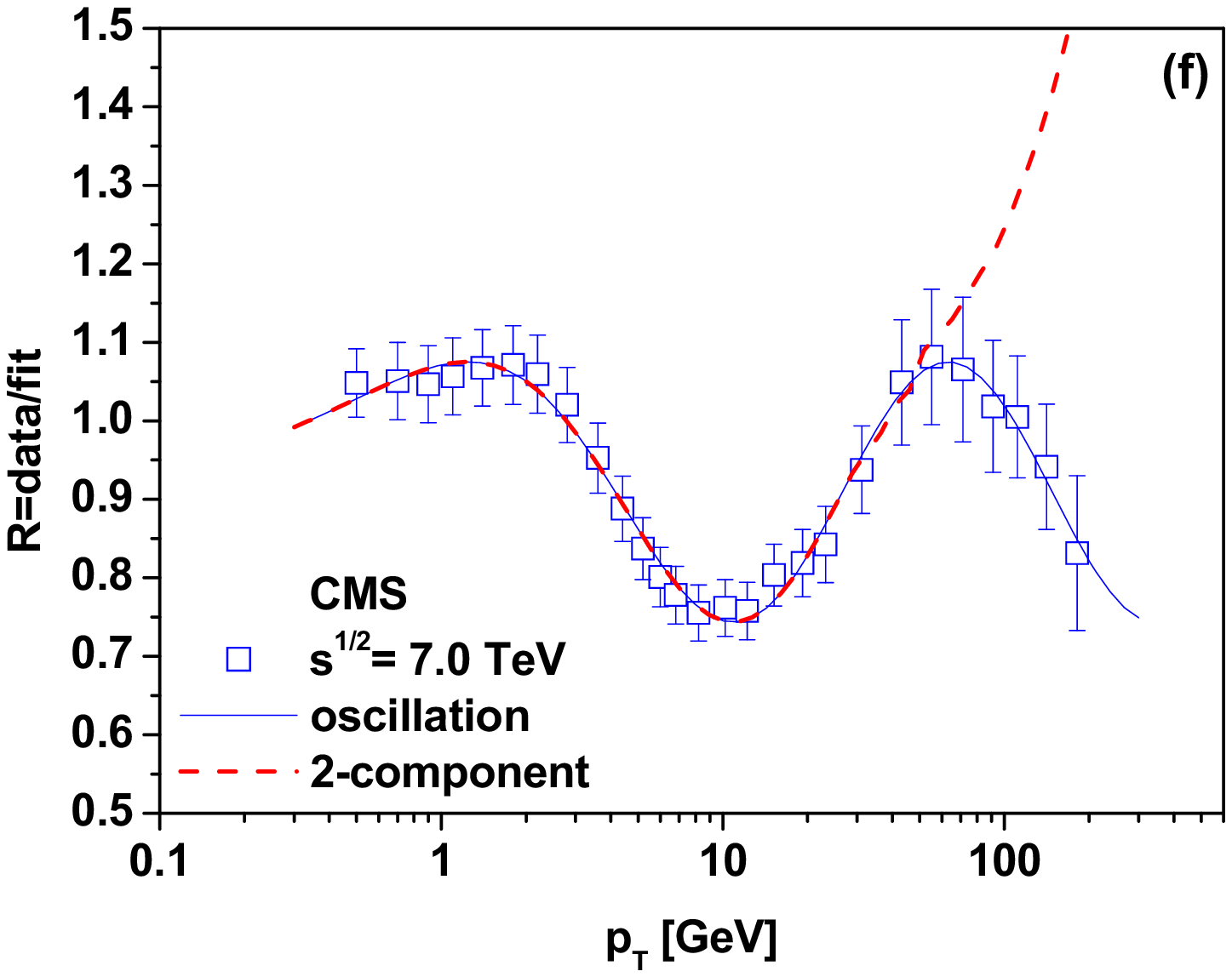}
}
\caption{(Color online) $(a)$ - Example of a typical $p+p$ large $p_T$ distribution observed in LHC experiments.
$(b)$ - Log periodically oscillating factor $R$ for different LHC experiments. $(c)$ and $(d)$ - The same factor $R$ at two different energies. $(e)$ - Example of a $Pb+Pb$ large $p_T$ cross section. $(f)$ - The $R$ factor for $Pb+Pb$ collisions; the prediction of the two-component approach proposed in \cite{SoftHard1,SoftHard2} is shown as the red line. For more details see text and \cite{WW_APPB,WW_ChSF}.}
\label{FF1}
\end{center}
\end{figure}

The first possibility may be related to scale invariance in the system \cite{Sornette} and was already widely discussed by us in \cite{WW_E,WW_APPB,WW_ChSF}. Such log-periodic oscillations are an immanent feature of any power-like distribution \cite{Sornette}. In \cite{WWln} we have shown that they also appear in quasi power-like distributions of the Tsallis type. In general they are attributed to a discrete scale invariance (connected with a possible fractal structure of the process under consideration) and are described by introducing a complex power index $n$, or $q$ in our case \cite{Sornette}. Such a case was discussed by us in \cite{WWln,RWWlnA,WW_E,WW_APPB,WW_ChSF} and has a number of observed consequences, like a complex heat capacity of the system, complex probability and complex multiplicative noise, all of them discussed briefly in \cite{WW_E}.

The other possibility is that the scale parameter $T$ has some specific log-periodic oscillations. To the best of our knowledge such a possibility has not so far been addressed. We discussed it in \cite{WW_APPB,WW_ChSF} where it was shown that the observed transverse momenta distributions presented in Fig. \ref{FF1} can be described using the following log-oscillating (as a function of transverse momentum $p_T$) temperature $T$ and keeping the power index $n$ constant and real:
\begin{equation}
T\left( p_T\right) = a + b \sin\left[ c \ln\left( p_T + d\right) + e \right]. \label{T_p_T}
\end{equation}
\begin{figure}[h]
\begin{center}
\resizebox{1.0\textwidth}{!}{%
  \includegraphics{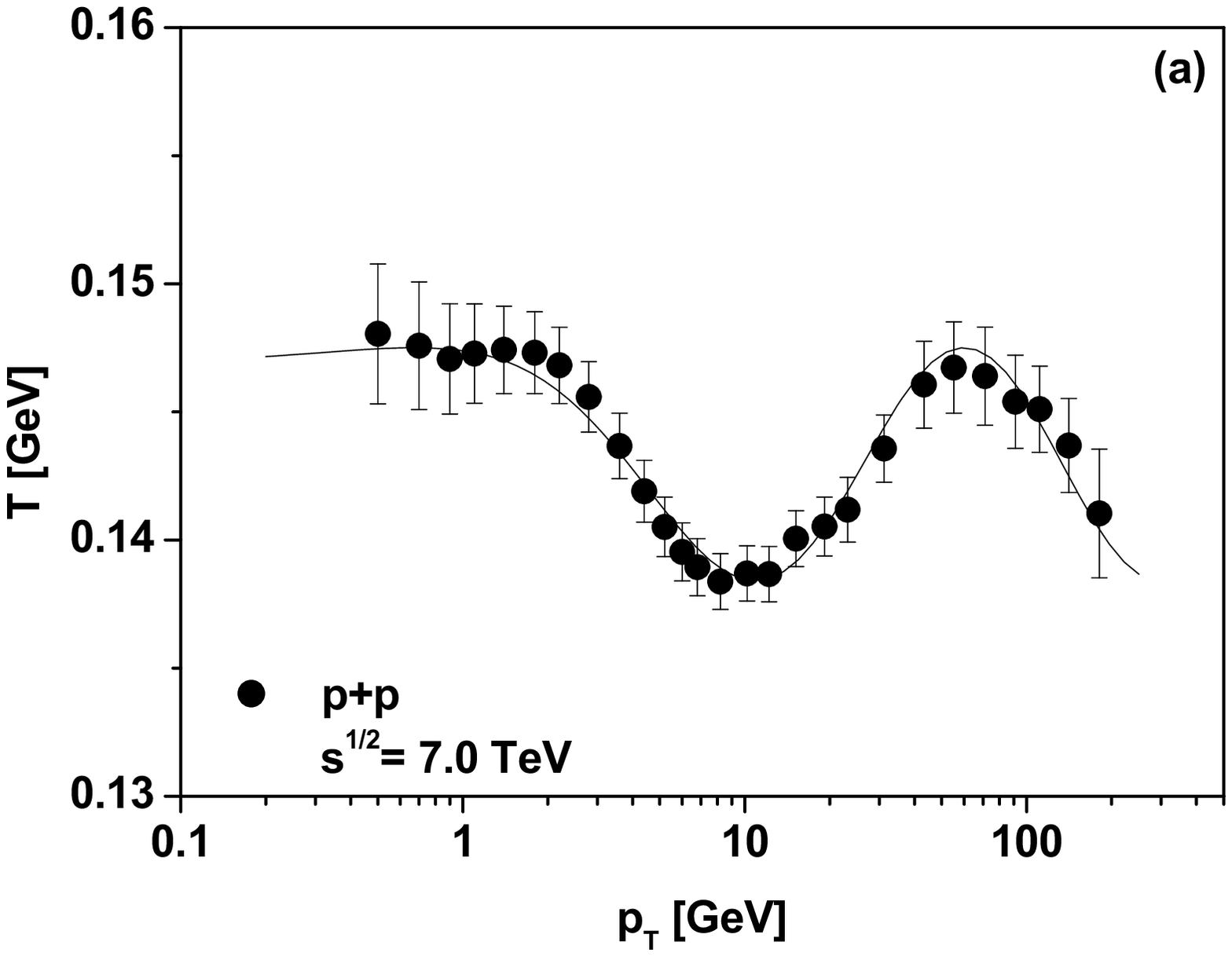}\hspace{5mm}
  \includegraphics{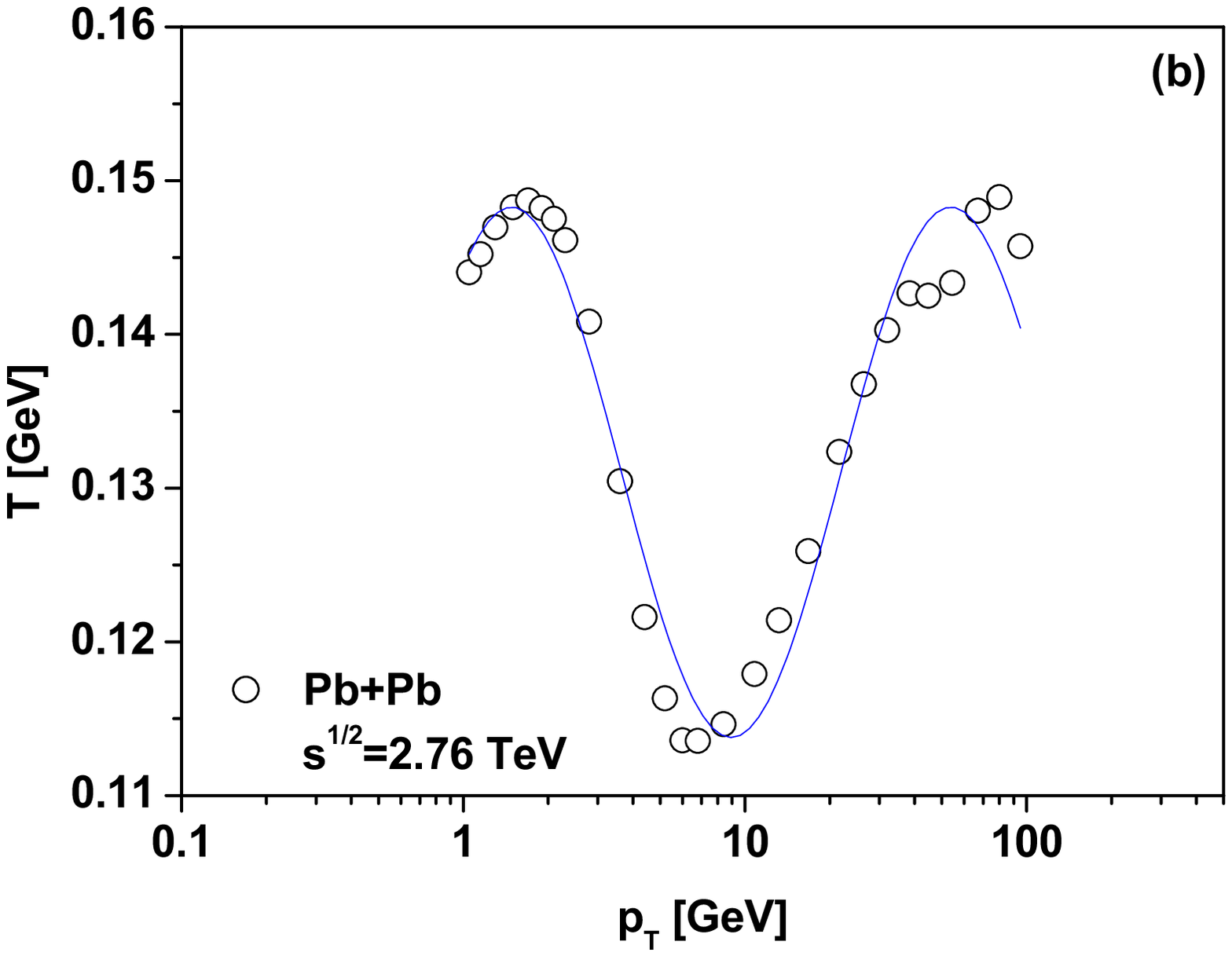}
  }
\caption{(Color online) Example of the observed log-periodicity of the scale parameter $T$ for $p+p$ (panel $(a)$) and $Pb+Pb$ (panel $b)$) collisions (deduced from ALICE data \cite{QP4,QP5}, see \cite{WW_APPB,WW_ChSF} for details). The fits follow Eq. (\ref{T_p_T}) with parameters $a = 0.143$, $b=0.0045$, $c=2.0$, $d=2.0$, $e=-0.4$ for $p+p$ collisions and $a=0.131$, $b=0.019$, $c=1.7$, $d=0.05$, $e=0.98$ for $Pb+Pb$ collisions.
}
\label{FF1a}
\end{center}
\end{figure}
It was shown there that the origin of such logarithmic oscillations (and, respectively, the meaning of the parameters $a$, $b$, $c$, $d$ and $e$ in Eq. (\ref{T_p_T})) comes from the possible energy dependence of the noise defining the corresponding stochastic processes in the well known stochastic equation for the evolution of temperature $T$
written in the Langevin formulation with $p_T$-dependent noise, $\xi\left(t, p_T\right)$, and allowing for time-dependent $p_T = p_T(t)$  \cite{Lang}:
\begin{equation}
\frac{dT}{dt} + \frac{1}{\tau} T + \xi\left( t,p_T\right) = \Phi. \label{LE}
\end{equation}
Here $\tau$ is the relaxation time, $\xi\left(t,p_T\right)$ is the $p_T$-dependent noise. Assuming additionally that we have time-dependent observed transverse momentum, $p_T = p_T(t)$, and that it increases following the scenario of the preferential growth of networks \cite{PrefGrowth} we get
\begin{equation}
\frac{dp_T}{dt} = \frac{1}{\tau_0}\left( \frac{p_T}{n} \pm T\right) \label{PrefG}
\end{equation}
(where $n$ coincides with the power index in Eq. (\ref{TD}); $\tau_0$ is some characteristic time step). The stationary dependence of $T\left( p_T\right)$ represented by Eq. (\ref{T_p_T}) is obtained in two cases (cf. Section 3.2 in \cite{WW_ChSF} or Section 5.2 in \cite{WW_APPB}): either the noise term increases logarithmically with transverse momentum while the relaxation time $\tau$ remains constant:
\begin{equation}
\xi\left(t, p_T\right) = \xi_0(t) + \frac{\omega^2}{n}\ln \left( p_T\right), \label{p_T_increase}
\end{equation}
or, equivalently, one keeps the $p_T$-independent white noise, $\xi\left( t, p_T\right) = \xi_0(t)$, constant but allows for the relaxation time becoming $p_T$-dependent, for example assuming that
\begin{equation}
\tau = \tau\left( p_T\right) = \frac{n\tau_0}{n + \omega^2 \ln\left( p_T\right)} \label{Tau_dep}
\end{equation}
(in both cases $\omega$ is some new parameter, cf. \cite{WW_APPB,WW_ChSF} for details). It turns out that to fit data one needs only a rather small admixture of the stochastic processes with noise depending on the transverse momentum (defined by the ratio $b/a \sim 3\%$). The main contribution comes from the usual energy-independent Gaussian white noise. It should be stressed that  whereas each of the proposed approaches is based on a different dynamical picture, nevertheless they are numerically equivalent.

\section{Oscillating $T$ as a signal of sound waves in hadronic matter}
\label{Fourier}

Whereas the occurrence of  a complex power index $n$ is a rather ubiquitous phenomenon in situations where one encounters scaling phenomena, a varying scale parameter (temperature) $T$ is not so widely known or used. However, the possible oscillatory  character of $T$ is not so exotic because it can be found in hydrodynamical models widely used to describe strongly interacting systems and also in astrophysics in the description of dense stars. Such models have recently become increasingly sophisticated and popular in response to the continuous support received from experiment. For example, a specially strong case is the observation of the so called  elliptic flow of secondaries produced in multiparticle production experiments. This phenomenon is not easy to explain in other approaches, which are more successful in the description of the distributions of the measured transverse momenta but use a fixed scale parameter $T$. Also, a number of observations strongly indicate that hadronic matter produced in heavy ion collisions at RHIC and the LHC behaves as a kind of perfect fluid. The observation of sound waves would therefore provide yet further support  for a hydrodynamical description of the multiparticle production process. Such waves could arise because in the initial phase of the collision process a number of highly energetic partons are created which subsequently lose their energy. This can proceed in two ways: either by exciting modes of the medium in {\it collisional energy loss}, or by radiating gluons in {\it radiative energy loss}. The further dissipation of the released energy resulting in thermalization depends on the character of the medium: in a {\it weakly coupled medium} it proceeds  through a cascade of collisions among quarks and gluons in the quark-gluon plasma, whereas in a {\it strongly coupled medium} the released energy is dissipated directly into thermal excitations and sound waves\footnote{Such phenomena were, in fact, already studied in \cite{HydroWaves}.}. Actually, in relativistic heavy ion collisions we may also have hard parton-parton collisions in which the outgoing partons have to traverse the surrounding fluid before escaping and forming jets which subsequently hadronize. Such partons may therefore form Mach shock waves during their passages, and this, in turn, will affect the transverse momentum distribution of the observed final particles.

In this work we present yet another possible consequence of the hydrodynamical picture of the production process, namely the possible formation and propagation of sound waves in hadronic matter. So far we have only shown that to fit data the temperature $T$ has to show log-periodic oscillations in transverse momentum, i.e., $T=T\left( p_T\right)$. To gain further insight into the collision process one should therefore form the Fourier transformation of such dependencies (as, for example, was done recently in \cite{Hindusi}). This will allow us to reconstruct the space-time structure of the collision process\footnote{Some explanatory remarks are in order here. In \cite{Hindusi} investigations were concentrated only on the possible fluctuations of $T$ whereas we also have at our disposal their additional log-periodic oscillations. The intrinsic fluctuations of $T$ investigated in \cite{Hindusi} are in our case accounted for by using  the Tsallis distribution which emerges from the usual Boltzmann distribution when one allows the scale parameter (temperature) to fluctuate according to a gamma distribution \cite{WWq}. Because in \cite{Hindusi} non-equilibrated systems were  described using the Boltzmann transport equation and their space variations were investigated by means of the Fourier transformations, the log-periodic oscillations were out of reach there.}.

To this end we investigate the use of the Fourier transform of the log-normal oscillations of $T\left( p_T\right)$ in the form presented in Eq. (\ref{T_p_T}):
\begin{equation}
T(r) = \sqrt{\frac{2}{\pi}}\int^{\infty}_0 \, T\left( p_T\right) e^{ip_T r} dp_T.   \label{FT}
\end{equation}
Because the oscillations are in transverse momentum $p_T$, $r$ is defined in the plane perpendicular to the collision axis and located at the collision point and denotes the distance from the collision axis.
\begin{figure}[t]
\begin{center}
\resizebox{1.0\textwidth}{!}{%
  \includegraphics{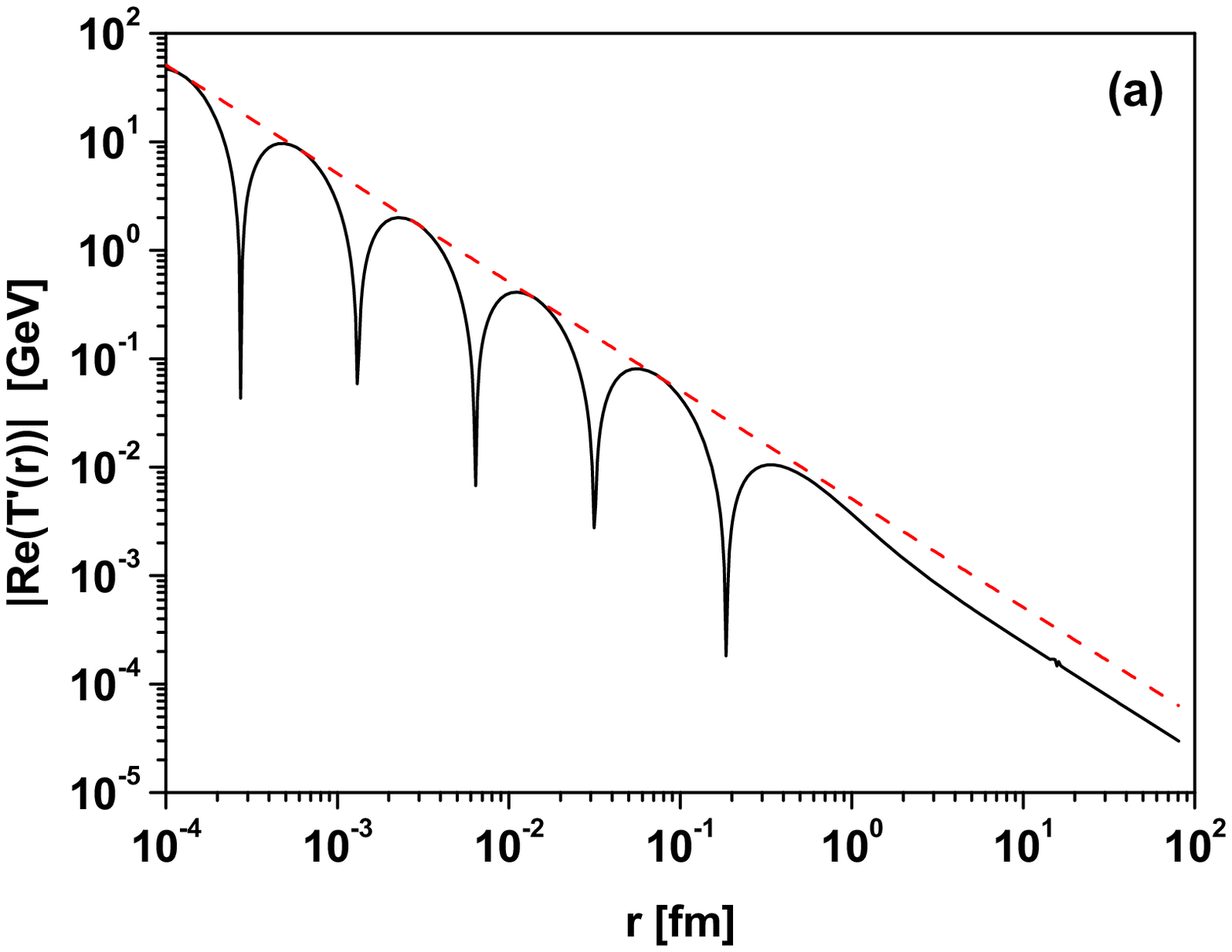}\hspace{5mm}
  \includegraphics{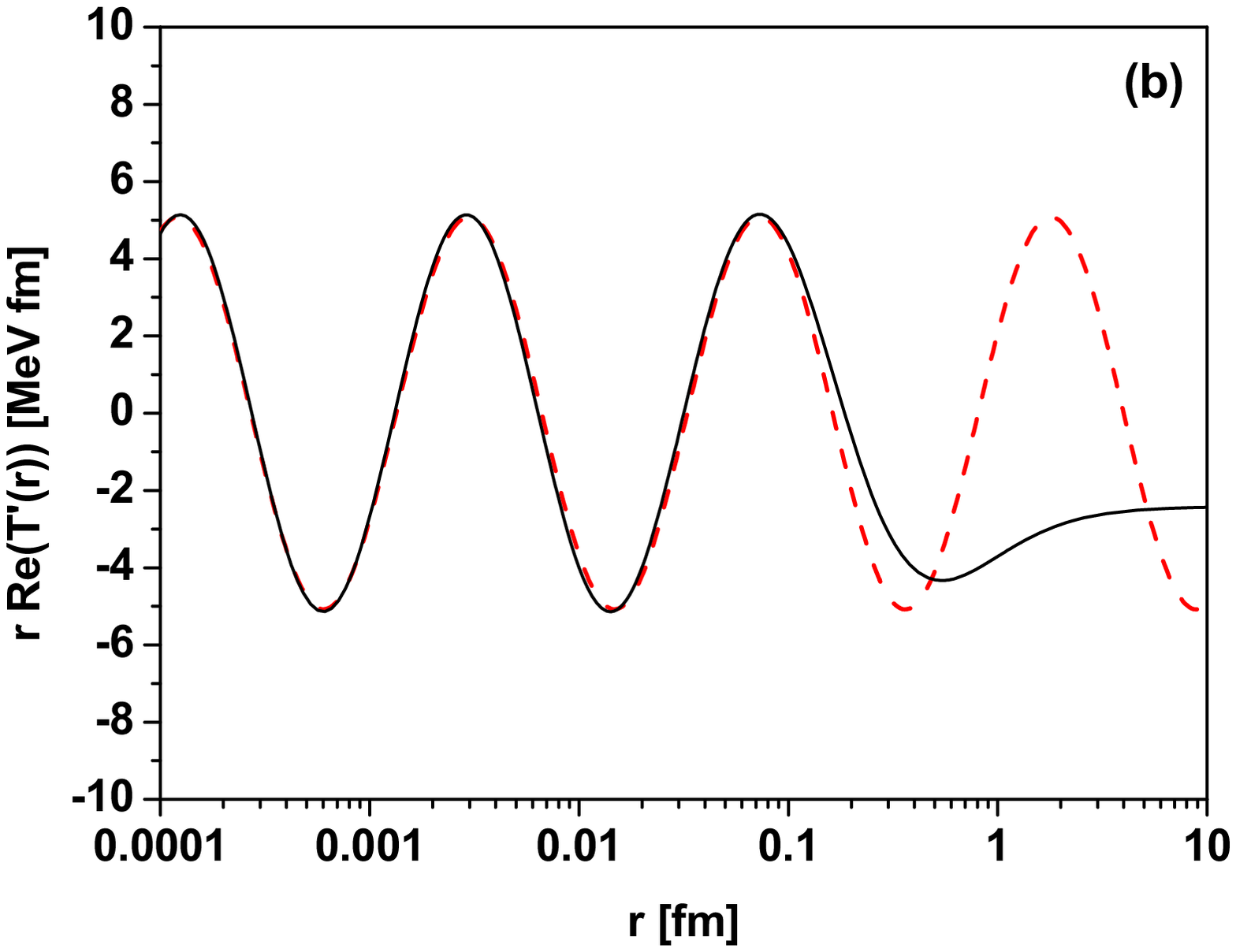}\hspace{5mm}
  \includegraphics{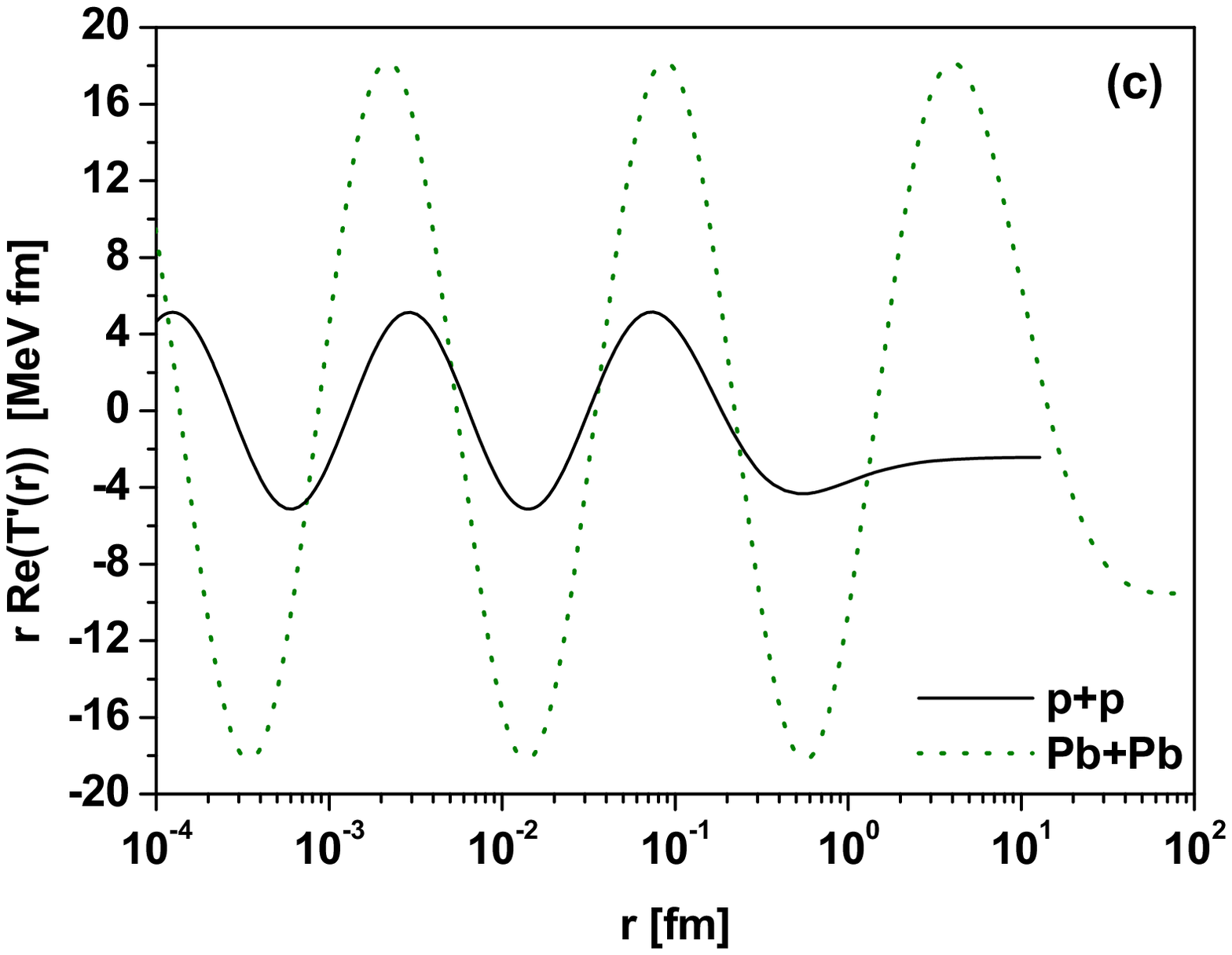}
  }
\caption{ (Color online) Panels $(a)$ and $(b)$:  The results of the Fourier transform of $T\left(p_T\right)$ from Eq. (\ref{T_p_T}) describing the results inferred from the CMS data \cite{QP1,QP2}. Panel $(c)$:  Oscillating behaviour of $r T'(r)$ for $p+p$ collisions at $7$ TeV presented in previous panels compared with similar results for the most central $Pb+Pb$ collisions at $2.76$ TeV (inferred from the ALICE data \cite{QP4,QP5}). See text for details.}
\label{FF1x}
\end{center}
\end{figure}
The results are shown in Fig. \ref{FF1x}. In panel $(a)$ we have $T'(r)$ (continuous line) as a function of $r$ confronted with $T'(r) = 0.0051/r$ dependence (dashed line).  In panel $(b)$ the continuous line represents $r T'(r)$ versus $r$ and  is confronted with the dashed line denoting the function $rT'(r) = 5.1 \sin [(2\pi)/3.2 \ln(1.24 r)]$ fitting it for small values of $r$. Panel $(c)$ presents a comparison of the oscillations of $T(r)$ for $p+p$ and the most central $Pb+Pb$ collisions deduced from the ALICE data \cite{QP4,QP5} at $2.76$ TeV \cite{RWWlnA} (the parameters of $T\left( p_T\right)$ used in both cases are the same as used in panel $(f)$ of Fig. \ref{FF1}. For the most central $Pb+Pb$ collisions the amplitude is $\sim 3.6$ bigger and the period of the oscillations is $\sim 1.15$ longer than in $pp$ collisions.  With decreasing centrality the amplitude in central $Pb+Pb$ collisions decreases smoothly reaching practically the same value as for $p+p$ collisions \cite{RWWlnA}. In  the case considered here, the region of regular oscillations can be fitted by
\begin{eqnarray}
\!\!\!\!\!(p+p):&& rT'(r) = 5.1 \sin \left[ \frac{2\pi}{3.2}\ln(r) + 0.42\right]; \label{R_Pb}\\
\!\!\!\!\!(Pb+Pb):&& rT'(r) = 18.53 \sin \left[ \frac{2\pi}{3.68}\ln(r) - 0.51\right]. \label{R_p}
\end{eqnarray}

To see how such a log-periodically oscillating structure of $T(r)$ occurs let us study the flow of a compressible fluid in a cylindrical source assuming small oscillations and oscillatory motion with small amplitude in a compressible fluid (i.e., a {\it sound wave}, at each point in the fluid it causes alternate compression and rarefaction of the matter). Since the oscillations are small, their velocity $v$ is also small and the term $( {\bold v}\cdot {\bold {grad}}){\bold v}$ in Euler's equation may be neglected. For the same reason, the relative changes in the fluid density and pressure are small and we may write \cite{Landau}
\begin{equation}
 P = P_0 + P',\quad\quad \rho = \rho_0 + \rho', \label{Prho}
\end{equation}
where $P_0$ and $\rho_0$ are the constant equilibrium density and pressure, and $P'$ and $\rho'$ are their variations in the sound wave $\left( \rho' << \rho,~~P' << P\right)$. Neglecting small quantities of the second order ($P'$, $\rho'$ and $v$ are regarded to be of the first order) the equation of continuity, $\partial \rho/\partial t + {\bold {div}}\cdot(\rho {\bold v}) =0$ becomes
\begin{equation}
\frac{\partial \rho'}{\partial t} + \rho_0 {\bold {div}}({\bold v}) = 0 \label{lindiv}
\end{equation}
and  Euler's equation reduces in this approximation to
\begin{equation}
\frac{\partial {\bold v}}{\partial t} +  \frac{1}{\rho_0}{\bold {grad}} P' = 0. \label{Euler}
\end{equation}
The linearized equations (\ref{lindiv}) and (\ref{Euler}) are applicable to the propagation of sound waves if $v<<c$ (where $c$ is the velocity of sound), which  means that $P' << P_0$. Note that a sound wave in an ideal fluid is adiabatic, therefore the small change $P'$ in the pressure is related to the small change $\rho'$ in the density by ($c$ is the velocity of sound)
\begin{equation}
P' = \left(\frac{\partial P}{\partial \rho_0} \right)_S \rho'  = c^2 \rho'\qquad {\rm where}\qquad c = \sqrt{\left( \frac{\partial P}{\partial \rho}\right)_S} .\label{PpRr}
\end{equation}
Substituting $\rho'$ from Eq. (\ref{PpRr}) into Eq. (\ref{lindiv}) one gets
\begin{equation}
\frac{\partial P'}{\partial t} + \rho_0 \left( \frac{\partial P}{\partial \rho_0}\right)_S  {\bold {div}}\cdot{\bold v} = 0 \label{Eq1}
\end{equation}
which, together with Eq. (\ref{Euler}), using the unknowns ${\bold v}$ and $P'$, provides a complete description of the sound wave we are looking for. To express all the unknowns in terms of one of them, it is convenient to introduce the velocity potential $f$ by putting ${\bold v} = {\bold {grad}} f$.  From Eq. (\ref{Euler}) we have the relation between  $P'$ and $f$:
\begin{equation}
P' = - \rho \frac{\partial f}{\partial t}. \label{Ppf}
\end{equation}
which, used together with Eq. (\ref{Eq1}), results in the following cylindrical wave equation which the potential $f$ must satisfy,
\begin{equation}
\frac{1}{r} \frac{\partial}{\partial r}\left( r \frac{\partial f}{\partial r}\right) - \frac{1}{c^2}\frac{\partial^2 f}{\partial t^2} =0.    \label{partialwe}
\end{equation}
It can be shown that this represents a travelling longitudinal plane sound wave with velocity ${\bold v}$ in the direction of propagation. It is related to the pressure $P'$ and the density $\rho'$ in a simple manner, namely
\begin{equation}
v = \frac{P'}{\rho c} = c \frac{\rho'}{\rho}. \label{vel}
\end{equation}
To relate the above results with the temperature note that we can write $T(r)$ as consisting of a constant term, $T_0$, and some oscillating addition, $T'(r)$:
\begin{equation}
 T(r) = T_0 + T'(r)\qquad {\rm where}\qquad T' = \left( \frac{\partial T}{\partial P}\right)_S P'.  \label{oscillatingT}
\end{equation}
Using the well known thermodynamic formula $\left( \partial T/\partial P\right)_S = \left(T/c_P\right)\left(\partial V/\partial T\right)_P$ and Eq. (\ref{vel}) one obtains that
\begin{equation}
T' = \frac{c \kappa T}{c_P} v \qquad {\rm where}\qquad \kappa = \frac{1}{V}\left( \frac{\partial V}{\partial T}\right)_P \label{ResLandau}
\end{equation}
where $\kappa$ is the coefficient of thermal expansion and $c_P$ denotes the specific heat  at constant presure  \cite{Landau}.
In the case of a monochromatic wave, when $f(r,t) = f(r)\exp(-i\omega t)$, we have that
\begin{equation}
\frac{\partial^2 f(r)}{\partial r^2} + \frac{1}{r} \frac{\partial f(r)}{\partial r} + K^2 f(r) = 0,\qquad \qquad
K=K(r) = \frac{\omega}{c(r)}  \label{eqn}
\end{equation}
where $K$ is the wave number which in inhomogeneous media depends on $r$. For the wave number given by
\begin{equation}
K(r) = \frac{\alpha}{r} \label{example}
\end{equation}
the solution of Eq. (\ref{eqn}) is given by a log-periodic oscillation in the form
\begin{equation}
f(r) \propto \sin [ \alpha \ln(r)]. \label{solution}
\end{equation}
Because ${\bold v} = {\bold {grad}} f$ we have $f(r) \propto v r$. Therefore, using Eq. (\ref{ResLandau}), we can write that
\begin{equation}
rT'(r) \propto \frac{c\kappa T_0}{c_P}f(r) = \frac{c \kappa T_0}{c_P} \sin[ \alpha \ln (r)]. \label{final}
\end{equation}
This is the solution we have used in describing the $T'(r)$ deduced from data and presented in Fig. \ref{FF1x}.

It should be mentioned at this point that the above problem can be considered from yet another point of view. Namely, it turns out that, according to \cite{BZ1,BZ2,B1,B2},  Eq. (\ref{eqn}) with the wave number (\ref{example}) has a so-called {\it self similar solution of the second kind}. Such a solution is known from other branches of physics and is connected with the so called {\it intermediate asymptotic} encountered whenever dependence on the initial conditions disappears (because sufficient time has already passed from the beginning of the process considered), but our system has not yet reached the state of equilibrium \cite{BZ1,BZ2,B1,B2}. Introducing the variable
\begin{equation}
\xi = \ln r \label{ksi}
\end{equation}
we find that Eq. (\ref{eqn}) for the wave number (\ref{example}) represents a travelling wave equation,
\begin{equation}
\frac{\partial^2 F(\xi)}{\partial \xi^2} + \alpha^2 F(\xi) = 0, \label{TWE}
\end{equation}
the solution of which is
\begin{equation}
F(\xi) \propto \sin(\alpha \xi). \label{TWS}
\end{equation}
Interestingly enough, this self similarity of Eq. (\ref{eqn}), which can be written as $F(\xi + \ln \lambda) = F(\xi)$, can be confronted with a kind of scale invariance of this equation, namely with the fact that $f(\lambda\cdot r) = f(r)$.

To summarize this part: The space picture of the collision (in the plane perpendicular to the collision axis and located at the collision point) presented in Fig. \ref{FF1x} (panels $(b)$ and $(c)$) shows us the existence of some regular (on the logarithmic scale) structure for small distances. For $p+p$ collisions it starts to weaken quite early (at $ r \sim 0.1$ fm) and essentially disappears when $r$ reaches the dimension of the nucleon, i.e., for $r \sim 1$ fm. For $Pb+Pb$ collisions it seems to last longer, to around $r \sim 10$ fm (i.e., to a typical dimension of the nucleus).

\section{Possible experimental consequences}
\label{Exp}

What can be deduced from our results? Note that the longer period of the oscillations in the $Pb+Pb$ collisions means that the values of the parameter $\alpha$ in Eq. (\ref{solution}) in nuclear collisions are smaller than those for $p+p$ collisions. Furthermore, considering the form of $K$ from Eq. (\ref{eqn}) or Eq. (\ref{example}), and remembering that $\omega/c(r)=\alpha/r$, one may deduce that the velocity of sound, $c(r) = (\omega/\alpha)r$, is greater in the nuclear environment  (for $Pb+Pb$) than in the $p+p$ case. This, in turn, means that the refractive index $n(r) = c_0/c(r)$ at position $r$ in nuclear collisions is smaller than in proton collisions. To summarize: In both cases we encounter an inhomogeneous medium with $r$-dependent  velocity of sound, $c(r)$, and refractive index, $n(r)$.

These findings seem to agree with the fact that in nuclear collisions one really observes a higher speed of sound as demonstrated by the NA61/SHINE collaboration at SPS energies \cite{NA61,GGS} (note, however, that what is measured is a parameter in the equation of state of hadronic matter described by a hydrodynamical model, $c^2_s$). This is not so unexpected because, considering the connection of the isothermal compressibility  of nuclear matter, $\kappa_T = - (1/V)\left( \partial V/\partial P\right)_T$, and fluctuations of the multiplicity of produced secondaries represented by the relative variance, $\varpi$, of multiplicity fluctuations, one finds that \cite{Hill,Balescu}
\begin{equation}
T\kappa_T \rho_0 = \frac{\langle N^2 \rangle - \langle N\rangle^2}{\langle N\rangle} = \varpi \label{IC-NF}
\end{equation}
(where $\rho_0 = \langle N\rangle/V$ denotes the equilibrium density for $N$ particles with mass $m$ located in volume $V$). This allows the velocity of sound to be expressed by fluctuations of multiplicity:
\begin{equation}
c = \sqrt{\frac{\gamma}{\kappa_T \rho_0 m}} = \sqrt{\frac{\gamma T}{\varpi m}}\quad {\rm or}\quad \varpi = \frac{\gamma T}{m}\cdot \frac{1}{c^2}\quad{\rm where}\quad \gamma = \frac{c_P}{c_V}. \label{SV}
\end{equation}
Note that a higher velocity of sound $c$ corresponds to lower fluctuations of multiplicity $\varpi$. From the experimental data shown in Fig. \ref{Fig4} \cite{NA61,GGS,RecentNA61} one has that
\begin{equation}
\frac{c_{Pb+Pb}}{c_{p+p}} \simeq 1.04\quad{\rm and}\quad \frac{\varpi_{p+p}}{\varpi_{Pb+Pb}} \simeq 1.29 \pm 0.04 \label{results}
\end{equation}
\begin{figure}[h]
\begin{center}
\resizebox{0.55\textwidth}{!}{%
  \includegraphics{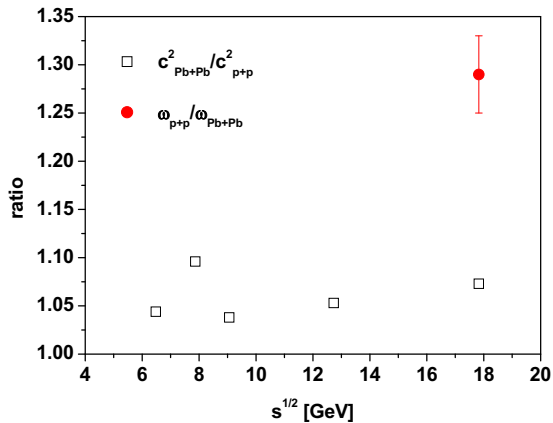}
}
\caption{(Color online) The ratio $c^2_{Pb+Pb}/c^2_{p+p}$ presented in \cite{NA61,GGS} compared with the ratio $\omega_{p+p}/\omega_{Pb+Pb}$ taken from \cite{RecentNA61}. }
\label{Fig4}
\end{center}
\end{figure}
From them, using Eq. (\ref{SV}), one can deduce the expected ratio of the factors $\gamma T$ for $p+p$ and $Pb+Pb$ collisions at a beam energy of $158$ GeV:
\begin{equation}
\frac{\left(\gamma T\right)_{p+p}}{\left( \gamma T\right)_{Pb+Pb}} \sim 1.2. \label{ratio}
\end{equation}
This could be the subject of further experimental investigations.

We close with the remark that these results may be connected with the pair correlation function, $g^{(2)}$, because the scaled variance can be written as \cite{FS}
\begin{equation}
\rho_0 \kappa_T T = 1 + \rho_0 \int d \vec{r} \left[ g^{(2)}(r) - 1\right]. \label{Corrrel_g}
\end{equation}
As shown in \cite{RW}, for central nuclear collisions the number of binary collisions exceeds that of wounded nucleons (each nucleon participates in a number of collisions with other nucleons). This results in the correlation function becoming negative which, in turn, leads to a diminishing of fluctuations of multiplicity (because the variance of the total multiplicity from a number of particular collisions is smaller that the sum of the variances of independent nucleon-nucleon collisions).

\section{Summary}
\label{Sum}

Data from all LHC experiments strongly suggest that transverse momentum distributions are not only described by a quasi-power law (Tsallis distribution with power index $n$ and scale parameter $T$ usually identified with temperature) but also that these distributions are additionally decorated with some characteristic log-periodic oscillations clearly visible for large $p_T$ events. Whereas this is a rather subtle effect, it is very persistent. It shows itself in all experiments, at all energies (provided that the range of transverse momenta observed is large enough) and also in reactions with nuclei where they grow with increasing centrality of the collision, and finally, they cannot be erased by any reasonable change of fitting parameters. All these points strongly suggest that these oscillations deserve to be taken seriously as a real effect which should be studied in detail.

Such studies reveal that either the system and/or the underlying physical mechanisms have characteristic {\it scale invariance behavior} (resulting in the power index $n$ becoming complex) or that we observe a {\it sound wave in hadronic matter} (resulting in the temperature oscillations)  which has a self similar solution (in log-periodic form). In the former case the discrete scale invariance and its associated complex exponents $n$ can appear spontaneously, without a pre-existing hierarchical structure \cite{Sornette}. In the latter case the corresponding wave equation has {\it self-similar solutions of the second kind} connected with the so called {\it intermediate asymptotic} (observed in phenomena which do not depend on the initial conditions because sufficient time has already passed, although the system considered is still out of equilibrium) \cite{BZ1,BZ2,B1,B2}. This suggests that both in $p+p$ and $Pb+Pb$ collisions one deals with an inhomogeneous medium with the density and the velocity of sound both depending on the position and this can have some interesting experimental consequences.

\medskip
This research  was supported in part (GW) by the National Science Center (NCN) under contract 2016/22/M/ST2/00176.  We would like to thank warmly Dr Nicholas Keeley for reading the manuscript.

\end{document}